\documentclass[11pt]{article}
\pdfoutput=1
\usepackage{times}
\usepackage{geometry}
\usepackage[utf8]{inputenc}
\usepackage{enumitem,amssymb}
\usepackage{amsmath}
\usepackage{graphicx}
\usepackage{fancyhdr}
\usepackage{aas_macros}
\usepackage{mdframed} 
\usepackage{color}
\usepackage{pdfpages}
\usepackage{url}

\usepackage[authoryear]{natbib}
\setcitestyle{authoryear,open={(},close={)}}

\mdfdefinestyle{theoremstyle}{
innertopmargin=\topskip,}
\mdtheorem[style=theoremstyle]{lrptextbox}{}

\usepackage{xcolor}

\begin{document}

\begin{center}
\textbf{\LARGE{Opportunities and Outcomes for Postdocs in Canada}}
\end{center}

\noindent {\bf Authors:} Henry Ngo* (NRC Herzberg), Helen Kirk (NRC Herzberg), Toby Brown (McMaster), Tyrone E. Woods (NRC Herzberg), Gwendolyn Eadie (U. Toronto), Samantha Lawler (Campion College/U. Regina), Locke Spencer (U. Lethbridge)\\
\noindent * Corresponding author contact info: henry@planetngo.ca\\

\noindent {\bf Topic areas:} state of the profession; training, careers, demographics and professional development; equity, diversity and inclusion\\

\noindent This version is modified from the official submission in order to include the full appendix. The original submission can be found at \url{http://bit.ly/LRP2020_white_papers}. The associated Expression of Interest ID for this paper is E028. The associated webpage for this paper is \url{http://rebrand.ly/lrp2020pdf}, where members of the Canadian astronomical community can sign their name in support of the ideas and recommendations presented here.

\vspace{1cm}

\section*{Introduction}

The postdoctoral fellow (PDF) career stage is, by design, transient and insecure. The short term nature of PDF employment means PDFs spend much of their time in their current position applying for their next one. The scarcity of continuing/permanent research positions, along with the precarious employment status and not knowing where the next position may be located contribute to a lot of stress for PDFs. These potentially long distance and regularly occurring moves can cause financial strain on PDFs\footnote{Kwok (2019), Nature, 571, 135}. PDF stress is well known---career sections of Science magazine contains articles with advice on managing stress at the PDF level\footnote{\url{https://www.sciencemag.org/careers/2014/07/stressed-out-postdoc}}. The Canadian Association of Postdoctoral Scholars' 2019 Pre-budget Briefing warned that low compensation and insecure employment leads to loss of research talent as PDFs leave their research field or seek positions outside Canada\footnote{\url{http://www.caps-acsp.ca/wp-content/uploads/2018/09/CAPS-2019-PreBudget-Brief-_-FINA-_-for-posting.pdf}}. Ultimately, there is no doubt that the combination of these factors has led to an increase in the transient scientific workforce, with the half-life of an astronomy career\footnote{Time for 50\% of a cohort to leave the field, as defined in Milojević et al. (2018)}  dropping from 37 years in the 1960s to only 5 years in 2007\footnote{Milojević et al. (2018) PNAS, 115, 12616}.

Indeed, these issues have been well documented, but for a first-hand account, the authors invite the reader to buy their friendly neighbourhood postdoc a coffee and listen to their story. These same sentiments were also reported in the LRP2020 PDF/RA town hall held in May 2019.

The authors of this report are a group of current and former PDFs in Canada. To collect data for this report, we conducted several surveys of PDFs and PDF employers in Canada. In the following sections, we present our view of the major issues affecting Canadian PDF opportunities and outcomes. We divide these issues into three calls to action, with each demonstrating a theme where improvement is needed. These themes relate to employment conditions for PDFs, research funding for PDFs and problems with the current way PDFs are hired in Canada. For each theme, we address the relevant issues and make recommendations for consideration by the Canadian astronomy community. 

We provide online supplementary material to this report at \url{http://rebrand.ly/lrp2020pdf}\footnote{At time of writing, this link redirects to \url{http://planetngo.ca/lrp2020_pdf_wp/}}, where the reader can find additional data, a list of signatures of Canadian astronomy community members in support of this report and a page where those interested can add their own signature.

\subsection*{Summary of Recommendations}

We list all of our recommendations below. In each of the following sections, we also discuss how each recommendation is relevant to the themes of that section.

\begin{flushleft}
\begin{itemize}
  \item[] {\bf R1. PDFs should be hired and compensated as skilled experts in their areas, not as trainees.} This recommendation is discussed in all three sections.
  \item[] {\bf R2. Standard PDF hiring practices should be revised to be more inclusive of different life circumstances.} This recommendation is discussed in all three sections. Sub-items 2.1, 2.2 and 2.3 are discussed in Sections 1, 2, and 3; respectively.
    \begin{itemize}
        \item[] \textbf{R2.1} Allow PDFs the option of part-time employment.
        \item[] \textbf{R2.2} Remove years-since-PhD time limits from PDF jobs.
        \item[] \textbf{R2.3} Financially support PDF hires for relocation and visa expenses.
    \end{itemize}
  \item[] {\bf R3. CASCA should form a committee to advocate for and provide support to astronomy PDFs in Canada.} This recommendation is discussed in Section 1.
  \item[] {\bf R4. CASCA should encourage universities to create offices dedicated to their PDFs.} This recommendation is discussed in Section 1.
  \item[] {\bf R5. PDFs and other PhD-holding term researchers with a host institution should be able to compete for and win grants to self-fund their own research.} This recommendation is discussed in Section 2.
  \item[] {\bf R6. Astronomy in Canada should hire general-purpose continuing support scientist positions instead of term PDFs to fill project or mission-specific requirements.} This recommendation is discussed in Section 3.
\end{itemize}
\end{flushleft}

\section{Our community needs to provide better terms of employment and benefits for our postdocs to all thrive and excel at their work.}\label{sec:support}

Hiring a PDF researcher on an individual research grant, such as an NSERC Discovery Grant, can be challenging, thus many of the PDF hires are made through larger programs.  We surveyed a number of programs across Canada which consistently hire PDFs regarding their provided benefits and programmatic issues.  These results are summarized in Table~1, while the full long-form responses to survey questions are provided in Table~2\footnote{Table~2 was not included in the official submission but can be found in the appendix.}.  A brief summary of comparable features offered by the United States' Hubble Fellowship is also included.

Several aspects are immediately noticeable from our benefits survey. First, the highly-competitive NSERC-funded PDF positions (NSERC PDF and Banting Fellowship) do not guarantee many of the basic benefits offered by the other programs, both in the personal (e.g., extended health and dental coverage) and in the research (e.g., conference travel funds) domains.  Despite the high degree of competitiveness for these positions, unless additional resources are provided by the host institution, these PDFs work under a significant disadvantage compared to their peers hired for other Canadian named fellowships.  This is especially true for the NSERC PDF, where the base salary is \$15,000 to \$20,000 lower than most of the other major fellowships in Canada.

Secondly, many of the basic benefits that non-academic employees have (e.g., parental leave, paid vacation and sick leave) are almost always offered at some level.  This is in contrast to the Hubble Fellowship, which lacks many of these provisions. Additionally, all of the Canadian fellowships surveyed allow for a fellowship extension following paid parental leave, unlike the Hubble Fellowship.

A third notable aspect is the level of variation between the fellowships in both personal benefits and research support.  For example, there is a range of more than \$8,000 per year in salary and \$10,000 per year in research expenses.  Types of paid leave in addition to sick leave and vacation leave vary widely between fellowship, and the level of paid top-up funding to EI for maternity and parental leave also vary significantly, from several weeks to nearly a full year at more than 90\% of full salary. 

\subsection*{Recommendations}
\vspace{-0.5cm}
\begin{flushleft}
\fbox{%
    \parbox{\textwidth}{%
        \bf R1. PDFs should be hired and compensated as skilled experts in their areas, not as trainees.}%
}
\end{flushleft}
It is well known that in today's employment landscape, many PDFs do not obtain permanent positions in astronomy. We therefore believe that it is important for PDFs to be treated as much like an early career professional in any other field of work, and not thought of as ``toughing it out'' until they get their dream faculty job.  In other words, fair compensation and benefits for the duration of the fellowship should be offered.

Our survey of the major fellowships available across Canada can provide guidance for best practices and areas of potential improvement for both individual PI and institutional PDF policies. Salary and research funding for the PDF are two examples which may prove challenging due to the nature of the Canadian funding landscape.  On the other hand, some other benefits, such as paid leave for surgery / hospitalization or compassionate leave, will be used only rarely, but may make an enormous difference for those few who need it.  Where possible, expanding benefits such as paid leave, relocation expenses, and paid parental leave, will create a more supportive environment for all PDFs, and may be particularly helpful for retention of PDFs from underrepresented groups\footnote{See, for example, {\tt https://www.nature.com/magazine-assets/d41586-019-02047-z/d41586-019-02047-z.pdf}}.
\vspace{-0.5cm}
\begin{flushleft}
\fbox{%
    \parbox{\textwidth}{%
        {\bf R2. Standard PDF hiring practices should be revised to be more inclusive of different life circumstances.}
        
        \textbf{R2.1} Allow PDFs the option of part-time employment.}%
       
}
\end{flushleft}
While funding timescales may sometimes present challenges for implementation, allowing part-time PDFs is another direction that could help with retaining a broader spectrum of astronomers, e.g., parents with young children or family caregivers. As a personal example, one of the co-authors of this paper (SML) worked in a PDF position at half-time employment for one year because of family commitments. Having this as an option allowed SML to achieve the work-life balance she needed at that stage in parenthood, while she remained active in the field of astronomy.  She returned to full-time employment for a second PDF, and is now tenure-track faculty in astronomy in Canada.  She believes that this period of half-time employment is what kept her in the field, instead of ``leaking out of the pipeline'' like a disproportionately large fraction of women do during PDF years\footnote{Flaherty (2018), arXiv:1810.01511}.

Our PDF benefits survey was necessarily limited to major fellowships across Canada.  PDFs hired by individual PI grants may face entirely different circumstances than what our survey captured.  We therefore recommend that CASCA surveys all Canadian PDFs to determine working conditions such as salaries, benefits, contract lengths, and experience inside and outside of Canada.  Only once the entire PDF employment landscape is understood can informed decisions be made to ensure fair treatment and improved retention of underrepresented groups.
\vspace{-0.5cm}
\begin{flushleft}
\fbox{%
    \parbox{\textwidth}{%
        \bf R3. CASCA should form a committee to advocate for and provide support to astronomy PDFs in Canada.}%
}
\end{flushleft}
At the time of writing of this report, some of the authors have submitted a proposal for a Postdoc Advocacy Committee to the CASCA Board. Ideally, this committee would include both current PDFs as well as early career researchers in permanent positions to help ensure continuity of committee membership.  This committee would be well-placed to undertake the PDF survey mentioned above and other important initiatives.
\vspace{-0.5cm}
\begin{flushleft}
\fbox{%
    \parbox{\textwidth}{%
        \bf R4. CASCA should encourage universities to create offices dedicated to their PDFs.}
    }%
\end{flushleft}
We note that some universities, such as McMaster, have a dedicated Office of Postdoctoral Affairs.  A university-wide office dedicated to PDFs can help to ensure fair hiring practices and can provide targeted support and training to a group that often falls outside of the typical university categories of students and faculty. For example, the McMaster office co-sponsors career skills development sessions offered by MITACS and directs advertising to the PDF community.  

\section{Our community needs a mechanism for non-permanent research staff to compete for and win grants}

At the moment, there are no ``soft-money'' grants available to PDFs working in Canada, regardless of their demonstrable ability to carry out self-guided research. Those grants which are available (through, e.g., CSA, NSERC, etc.) generally provide insufficient funding to support a PDF's typical salary and benefits on their own, and regardless require one already hold a faculty position or equivalent in order to be an eligible PI. This is in stark contrast with the funding environment in the United States, where numerous opportunities exist for non-tenured scientists to apply for funding not only to support their research but to cover their own costs as well. At the moment, the absence of such funding in Canada creates the worrying potential for a ``funding gap'' in the career progression available to Canadian astronomers, particularly given the combination of a very long average time between obtaining a PhD and finding a faculty position\footnote{Metcalfe T.~S., 2008, PASP, 120, 229}$^{,}$\footnote{Perley D.~A., 2019, PASP, 131, 114502}, together with the arbitrary constraints on time post-PhD typical for most prize fellowships (see {\bf R2.2} below). This lack of funding leads to many facing a stark choice between leaving astronomy, or leaving Canada (see survey results in appendix), which in either case leads to a loss for the Canadian astronomical community. On the other hand, however, the ``patchwork'' nature of soft-money positions, with its accompanying absence of the benefits and security of contractually-backed employment, has been credited with producing an ever-growing cohort of ``second-class citizens'' within the US science community\footnote{Barinaga (2000), Science, 289:2024}, and contributing directly to difficulties in the retention of PDFs from under-represented groups (see previous section).

How then can we as a community provide a stable, equitable, sensible path for academics to grow from PhD to permanence, including adequate funding to foster the independent research leadership that will be essential in their roles as faculty? An alternative, albeit more difficult to implement, approach may be a more formalized ``ladder'' of fellowships, from early career research to faculty, each with appropriate accompanying funding provided to support independently-led research projects, and each with University-backed benefits. This has been successfully implemented abroad (e.g., Veni/Vidi/Vici in the Netherlands, DECRA/Future Fellowship in Australia, see also the EU's ECR scheme), although, at least anecdotally, there is some suggestion that these schemes may eventually suffer from ``credential creep,'' with it becoming increasingly difficult to leverage the highest awards within each scheme for a permanent position. Ultimately, however, this is likely tied to the extreme over-saturation of the available pool of PhD graduates relative to the permanent job market in astronomy\footnote{See, e.g., Cooray A., Abate A., H{\"a}u{\ss}ler B., Trump J.~R., Williams C.~C., 2015, arXiv, arXiv:1512.02223}, and may not indicate a problem with fellowship ladders themselves.

Implementing such an overhaul of the present funding structure would necessitate broad, interdisciplinary backing from the Canadian science community, likely requiring long-term advocacy. At present funding levels, this may also necessitate {\bf fewer} PDFs in Canadian astronomy. This can be offset, however, by a concurrent rise in the number of longer-term research staff hired by individual institutions to support the specific astronomy projects and facilities from which they benefit (see next section and \textbf{R6}). Creating a funding environment where non-faculty scientists can apply for research funding would be essential in order to allow staff in such long-term support positions to compete on an even footing and capitalize on their relative lack of administrative duties. This would be to the benefit of Canadian astronomy's research output as a whole, as is evident from the disproportionate role of non-faculty research staff at present in producing high-impact first-author publications (see Crabtree, LRP 2020 WP), and the modest hours available to regular faculty for conducting research\footnote{See, e.g., Koens, L.,R. Hofman \& J. de Jonge (2018). What motivates researchers? Research excellence is still a priority. The Hague: Rathenau Instituut}.

In the next decade, real progress can be made toward fostering a Canadian astronomical community where all early career researchers have the opportunity to pursue the funding necessary for ground-breaking, independently-led research, in an environment where they are provided with the competitive benefits considered standard for their skill-level. In the following, we outline several key practical steps toward this goal.

\subsection*{Recommendations}
\vspace{-0.5cm}
\begin{flushleft}
\fbox{%
    \parbox{\textwidth}{%
        \bf R1. PDFs should be hired and compensated as skilled experts in their areas, not as trainees.}%
}
\end{flushleft}
Treating PDFs as skilled experts means increasing salaries and benefits, bringing PDF jobs closer to the compensation provided by equivalent skill-level industry jobs or continuing positions at the same institution. We support these recommendations even if it means fewer 
PDF positions. Fewer, better compensated PDF positions will lead to less discrimination in who is financially able to accept a PDF, leading to fairly attracting highly talented researchers into Canadian PDF positions. This item is to be taken up by astronomers, particularly senior ones responsible for hiring PDFs through their grants or on fellowship committees.
\vspace{-0.5cm}
\begin{flushleft}
\fbox{%
    \parbox{\textwidth}{%
        {\bf R2. Standard PDF hiring practices should be revised to be more inclusive of different life circumstances.}
        
        \textbf{R2.2} Remove years-since-PhD time limits from PDF jobs.}%
       
}
\end{flushleft}
As discussed above, the length of time between PhD and a continuing astronomical research position is growing. Removing years-since-PhD time limits will prevent forcing people to ``leak'' out of the pipeline and faciliate the larger changes proposed here, where our community can support a larger population of non-faculty but continuing positions in research. We note that implementing \textbf{R2.2} in a vacuum may lead to undesirable consequences such as being hired as a PDF indefinitely. This recommendation should be considered alongside \textbf{R1} and \textbf{R6}. 
\vspace{-0.5cm}
\begin{flushleft}
\fbox{%
    \parbox{\textwidth}{%
        \bf R5. PDFs and other PhD-holding term researchers with a host institution (including longer-term research staff, see next section) should be able to compete for and win grants to self-fund their own research.}%
}
\end{flushleft}
This would require significant support and advocacy by LRPIC and CASCA to argue for funding agencies such as NSERC to change their policies. PDFs and term researchers would compete with faculty and other researchers with continuing positions.

\section{Our community needs to create a sustainable and mutually beneficial postdoc hiring model that considers the outcomes of their postdocs}

In LRP2010, PDFs were identified as ``critical contributor[s] to the success of astronomy'' and as ``mak[ing] significant contributions to research''. PDFs are an important part of the Canadian astronomical research ecosystem in ways beyond conducting research. In an informal survey of 29 PDFs, we found that PDFs also teach courses (24\%), mentor undergraduate students (90\%), mentor graduate students (72\%), serve on committees (55\%) and work with industry partners (14\%). Student mentoring work includes both research projects as well as career and professional development. Full survey results are presented in the appendix.

However, while our community has recognized the value of PDFs and LRP2010 even recommended an increase in Canadian PDFs, there is not enough support available for existing PDFs to thrive after they have finished their short term employment contracts. In some cases, well-intentioned support for PDFs may actually lead to worsening conditions. Here, we identify problems that PDFs encounter after their term ends. 

Finding the next academic position in Canada is difficult. While LRP2010 finds that Canada has a low ratio of PDFs (2 faculty per PDF instead of 1 to 1 as in other countries), there are still many more new PDF hires than continuing position hires (faculty or otherwise) in Canada. In our survey, 86\% and 35\% of PDFs indicate that they are interested in a research-track or teaching-track position at Canadian institutions, respectively, while there are fewer than 10 of these position types hired per year. In addition, there are arbitrary time restrictions placed on some PDF positions, with very few PDF options beyond 5 years past a PhD. This makes it difficult for some PDFs to find another PDF position in Canada. We also surveyed some PhD astronomers who left Canada for an international PDF position to find out reasons for leaving. Half of the respondents indicated they applied to PDF positions within Canada but there were not enough positions available so they did not receive an offer. 

These problems also affect PDFs during their term employment. The constant need to apply for jobs combined with a precarious employment situation and uncertainty in one's future create a lot of stress for PDFs. Moreover, the act of applying for jobs can be an arduous and time-consuming task that takes away from research. For PDFs who require a work permit, there are additional financial and time/effort costs to get a new work permit with every job change in Canada. And while many PDFs enjoy working with students, some types of arrangements, such as acting as an informal mentor without any official recognition, may take up a significant amount of time while not `counting' towards the training of highly-qualified personnel (HQP) valued by Canadian funding bodies\footnote{For instance, the Canadian Common CV, used by NSERC PDF, Banting PDF and NSERC Discovery Grant applications, specifically says that student supervision as part of PDF work does not count as HQP.} and may not be visible to future employers. In addition, while a PDF may also spend time mentoring students in areas of professional/career development (e.g. advise on grad school applications), this type of work isn't quantified at all in academic job and grant applications. 
Ultimately, all of these uncertainties negatively impact quality of life for PDFs and can interfere with their ability to fully thrive in the research environment.

There are many opportunities for astronomy PhD holders to find non-academic positions and indeed, 21\% of surveyed PDFs indicate they are seeking non-academic employment after their current PDF term. However, despite what LRP2010 says and what is commonly heard from the astronomical community, training as a PDF does not directly increase one's preparation for most non-academic career paths. In addition, since LRP2010, more and more universities are offering specialized degrees in data science or other related training programs, reducing the potential advantage of a science PhD in this career path.  Many astronomy-trained researchers that moved to other careers either received the majority of quantitative and technical skills needed in these other careers during graduate school or in their own time as a PDF. When considering job postings for non-academic careers and profiles of astronomers hired into non-academic careers, there appears to be little evidence that time worked as a PDF is inherently valuable for non-research careers. Perhaps time as a PDF can be helpful in increasing the overall years of work experience, but this can also been achieved outside of a PDF position, and likely be better compensated as well. Thus, while it is good that PDFs have these non-academic options, it is not accurate  to claim that a PDF position is beneficial for both academic \emph{and} non-academic career paths. \textbf{A PhD choosing a PDF position has made a commitment to the academic path and is paying real opportunity costs in doing so.} 

``More postdocs'' was a common refrain in LRP2010 town halls, and the report itself called for an increase in PDFs to support missions and facilities. In LRP discussions at the 2019 CASCA AGM, it was further argued that a steady flux of new PDFs is needed to keep up with the changing skill requirements of new projects. This position, however, contradicts the simultaneously presented argument that PDF experience is good preparation for jobs with nominally related skills in the private sector. How is it not possible for a PDF to develop new skills for a different astronomical project, yet entirely feasible to find a career outside of astronomy entirely? 

Here, we caution against an increase in PDFs without considering opportunities available for PDFs after the end of their term. We make recommendations below to avoid a future where our community hires researchers in term positions in order to increase overall research output in an exploitable and unsustainable manner. 

\subsection*{Recommendations}
We assert that PDFs be considered early career researchers instead of ``trainees'' as implied by some language in LRP2010. We hope our community creates a PDF employment model that provides mutual benefits for PDFs and their employers. The following recommendations are in support of these goals.
\vspace{-0.5cm}
\begin{flushleft}
\fbox{%
    \parbox{\textwidth}{%
        \bf R1. PDFs should be hired and compensated as skilled experts in their areas, not as trainees.}%
}
\end{flushleft}
As with any new position, while a new PDF may still learn some skills specific to their position (e.g. a new software or student supervision), PDFs have already completed their main training in graduate school. While there is a need for term positions in academia, the salary and benefits of these positions should still be commensurate with the PhD's experience and should reflect the employment of an expert, not a trainee. PhDs are treated as experts in non-academic career paths, where they are hired into senior analysis positions owing to the large number of years of training and experience required to do the original research necessary for a PhD.  PhDs in non-science fields are also treated as senior experts and are hired directly into faculty positions.

In addition, our community should hire and compensate term employees such as PDFs in the same manner as continuing staff. While the nature of academic research work drives a need for limited term positions to address specific needs within time-limited projects, it is not ethical for our community to inadvertently create poor employment conditions due to the nature of term positions.  

In particular, we are concerned that term employment allows employers to continually hire PDFs and new PhDs in revolving-door style without also hiring continuing positions. We are also concerned that term employment leads to PDFs working with lower salaries and fewer benefits compared to continuing positions outside of academia for someone with similar training and experience (see Tables~1 and 2). To address this, PDF employers should hire PDFs on term employment contracts that mirror those of their continuing research staff and/or faculty but at a rank that reflects the PDF's level of experience. We point to NRC Herzberg as an example, as PDFs at NRC Herzberg enjoy the same compensation and benefits as the continuing staff.
\vspace{-0.5cm}
\begin{flushleft}
\fbox{%
    \parbox{\textwidth}{%
        {\bf R2. Standard PDF hiring practices should be revised to be more inclusive of different life circumstances.}} 
}
\end{flushleft}

Our community should ensure our PDF employment practices do not create equity and inclusivity issues. Currently, PDFs are paid less than equivalently skilled workers outside of academia and their employment often requires relocation. Both of these issues create barriers to PDFs that might otherwise like to stay in academia. Because the only academic career options in Canadian astronomy --- aside from Faculty positions which are few and far between --- are another PDF or a continuing research or teaching position, precarious job security may create a selection effect on who is able to finally secure a continuing position in astronomy. We can improve our ability to attract and retain talented researchers by ensuring we are not selecting against people who need a salary commensurate with their ability and people who are not able to move around the country. 
\vspace{-0.5cm}
\begin{flushleft}
\fbox{%
    \parbox{\textwidth}{%
        \textbf{R2.3} Financially support PDF hires for relocation and visa expenses.}%
}
\end{flushleft}
Currently, PDFs are typically 1-3 years in length and may require a relocation that cost several thousand dollars to move within or to Canada. New PDFs will find it challenging to build these savings from graduate student stipends while current PDFs may have exhausted their savings with their last move completed 1-3 years ago. In addition, for non-Canadians, moving to work in Canada comes with considerable additional financial and time expenses to secure visas and work permits. Incorporating these benefits into the standard PDF hiring package will help open opportunities to more people. We note that for PDFs with partners, these frequent moves can lead to lost career opportunities and income for the partner as well. \vspace{-0.5cm}
\begin{flushleft}
\fbox{%
    \parbox{\textwidth}{%
        {\bf R6. Astronomy in Canada should hire general-purpose continuing support scientist positions instead of term PDFs to fill project or mission-specific requirements.}}
}
\end{flushleft}
Our community needs a new model for PDFs that not only harnesses their valuable research time and freedom for achieving research goals, but that also ensures they have fair and equitable working conditions which allow them to thrive both professionally and personally. Achieving this would require large structural changes in the way research is funded and conducted in Canada. One model to consider is for institutions to hire general purpose support scientists in continuing positions to fulfill the institution's project or mission-specific science duties. Although not permanent in the same way as tenured faculty, the incumbent would have severance benefits to help them find their next position, as is the case for many areas of employment without the concept of tenure.

One major challenge to these changes is whether there would be enough missions and projects to continuously hire people. Certainly, in a flat-funding scenario, this will raise the overall cost of doing research. We argue that it is preferable to have fewer well-supported PDF or PDF-like positions than to continue the existing model with only a slightly larger number of underpaid and precarious PDF positions. Continuing on the existing path may lead to a more exploitative job market and create a selection bias on researchers who are able to stay in academia.

\section*{Summary}
In this report, we present six recommendations to address three major areas of concerns affecting PDF employment and outcomes in Canada. Our main recommendations revolve around our community compensating PDFs as skilled experts instead of trainees (\textbf{R1}) and ensuring that PDF hiring practices are inclusive of the wide diversity of astronomers (\textbf{R2}). These two recommendations are part of the solution to all three issues. We also note that while a successful implementation of these recommendations would involve systemic changes to PDF employment in Canada, these recommendations are very much actionable at the level of an individual or committee responsible for PDF hiring decisions.

We also make two recommendations for CASCA, where we ask for a CASCA Postdoc Advocacy Committee (\textbf{R3}) and for CASCA to encourage universities to create offices dedicated to supporting PDFs at these institutions (\textbf{R4}). These leverage the wide network of our discipline's national society to pool resources and influence.

Finally, we make two recommendations intended to overhaul the employment of PDFs and early career researchers in Canadian Astronomy. \textbf{R5} asks for a major change in policies of funding agencies such as NSERC, which fund many disciplines outside of astronomy to allow researchers on term appointments to compete alongside tenure-track faculty for grant dollars. \textbf{R6} asks all astronomy institutions in Canada to end the practice of fulfilling their operational needs with serial term position hires instead of continuing positions. Whether these operational needs are to build a diverse research portfolio or to fulfill requirements for specific missions or projects, we advocate for hiring general-purpose astronomers that can modify their assigned work instead of hiring another term PDF without consideration of the outcomes for the outgoing PDF. These last two recommendations are ambitious for LRP2020 but we believe these ideas are important to consider as employment conditions evolve.

Reviewing these recommendations, we highlight that recommendations \textbf{R2.1}, \textbf{R3}, and \textbf{R4} come with no real costs and can be implemented immediately. \textbf{R2.2} requires coordination with \textbf{R1} and \textbf{R6}. We recognize that in a flat-funding scenario for Canadian astronomy research, there will be challenges for recommendations \textbf{R1}, \textbf{R2.3}, \textbf{R5}, and \textbf{R6}. However, we strongly believe this is the required path in order to sustain a successful, inclusive and ethical astronomy research program in the long term. With limited resources, we argue for a reduction in the number of PDFs in order to ensure each PDF can thrive. Although this may decrease the amount of science supported in Canada, we advocate for a solution that involves justifying additional funds instead of a solution that asks PDFs to accept precarious employment with little support. The latter (current) scenario creates a selection bias on the researchers themselves, which compromises the diversity of thought required for a successful national astronomy research program. In short, we ask for prioritization of people over production of papers.

\section*{Acknowledgements}
HK thanks all of the fellowship administrators who took the time to reply to the PDF benefits survey: Alice Chow (Dunlap), Carolina Cruz-Vinaccia (McGill), Candace Duong (CITA), Sheri Keffer (Perimeter), and Marie-France Nadeau (NRC). HN acknowledges that the main work site of his PDF position exists on the traditional territories of the Coast Salish peoples. Much of HN's contributions to this report took place while he was on paid Parental Leave, a benefit provided to Plaskett Fellows as they are NRC employees.

\clearpage
\section*{LRP text boxes}

\begin{lrptextbox}[How does the proposed initiative result in fundamental or transformational advances in our understanding of the Universe?]
A strong postdoctoral workforce enables fundamental and transformational advances. Implementing these recommendations will create a more inclusive and diverse research workforce, leading to a more diverse set of ideas, some of which may be transformational. In addition, ensuring opportunities are present for everyone will strengthen public support for astronomy in Canada.
\end{lrptextbox}

\begin{lrptextbox}[What are the main scientific risks and how will they be mitigated?]
N/A
\end{lrptextbox}

\begin{lrptextbox}[Is there the expectation of and capacity for Canadian scientific, technical or strategic leadership?] 
N/A
\end{lrptextbox}

\begin{lrptextbox}[Is there support from, involvement from, and coordination within the relevant Canadian community and more broadly?] 
Volunteers from the currently- and recently-PDF community in Canada wrote this report and solicited ideas from the same population. The Canadian astronomical community are encouraged to indicate their support for the ideas at \url{http://rebrand.ly/lrp2020pdf}
\end{lrptextbox}

\begin{lrptextbox}[Will this program position Canadian astronomy for future opportunities and returns in 2020-2030 or beyond 2030?] 
Yes, implementation of these recommendations will lead to a PDF community that is inclusive, thriving and poised to be Canada's astronomy leaders beyond 2030.
\end{lrptextbox}

\begin{lrptextbox}[In what ways is the cost-benefit ratio, including existing investments and future operating costs, favourable?] 
These recommendations aim to create fair and inclusive employment conditions for PDFs at the cost of the quantity of research time worked by the same PDFs. It is not reasonable to assign quantitative value to the former, however, as we put people ahead of papers, we consider the cost-benefit ratio to be favourable.
\end{lrptextbox}

\begin{lrptextbox}[What are the main programmatic risks
and how will they be mitigated?]
Some of our recommendations will increase the overall cost of doing science in Canada, and therefore we risk reducing the total scientific output without an increase in funding. This risk is mitigated by improving the overall quality of scientific work in Canada by creating inclusive opportunities for everyone to contribute novel ideas. One additional challenge is achieving buy-in support from astronomy departments at Canadian universities and institutions. We hope the LRP process and this report will raise awareness of the issue and the list of public supporters for this report will demonstrate the importance of these issues. The proposed CASCA Postdoc Advocacy Committee (Recommendation 3) will coordinate nation-wide efforts for a complete census of PDFs and from that, create coherent strategies to lead to nation-wide adoption. Our more ambitious recommendations involve changes at even higher levels, sometimes outside of astronomy itself. We acknowledge that this may not complete within the next decade but significant steps can be made.
\end{lrptextbox}

\begin{lrptextbox}[Does the proposed initiative offer specific tangible benefits to Canadians, including but not limited to interdisciplinary research, industry opportunities, HQP training,
EDI,
outreach or education?] 
These set of recommendations aim to improve opportunities, employment conditions and outcomes for PDFs in Canada. PDFs themselves are HQP. We aim for a more equitable, diverse and inclusive employment environment. The jobs created and supported by this environment will be open to Canadians. In addition, the products of PDF research in Canada will benefit Canadians generally, whether it is through additional education opportunities, outreach events and general sense of pride in one's own country and its scientific accomplishments.
\end{lrptextbox}

\includepdf[pages=1-last]{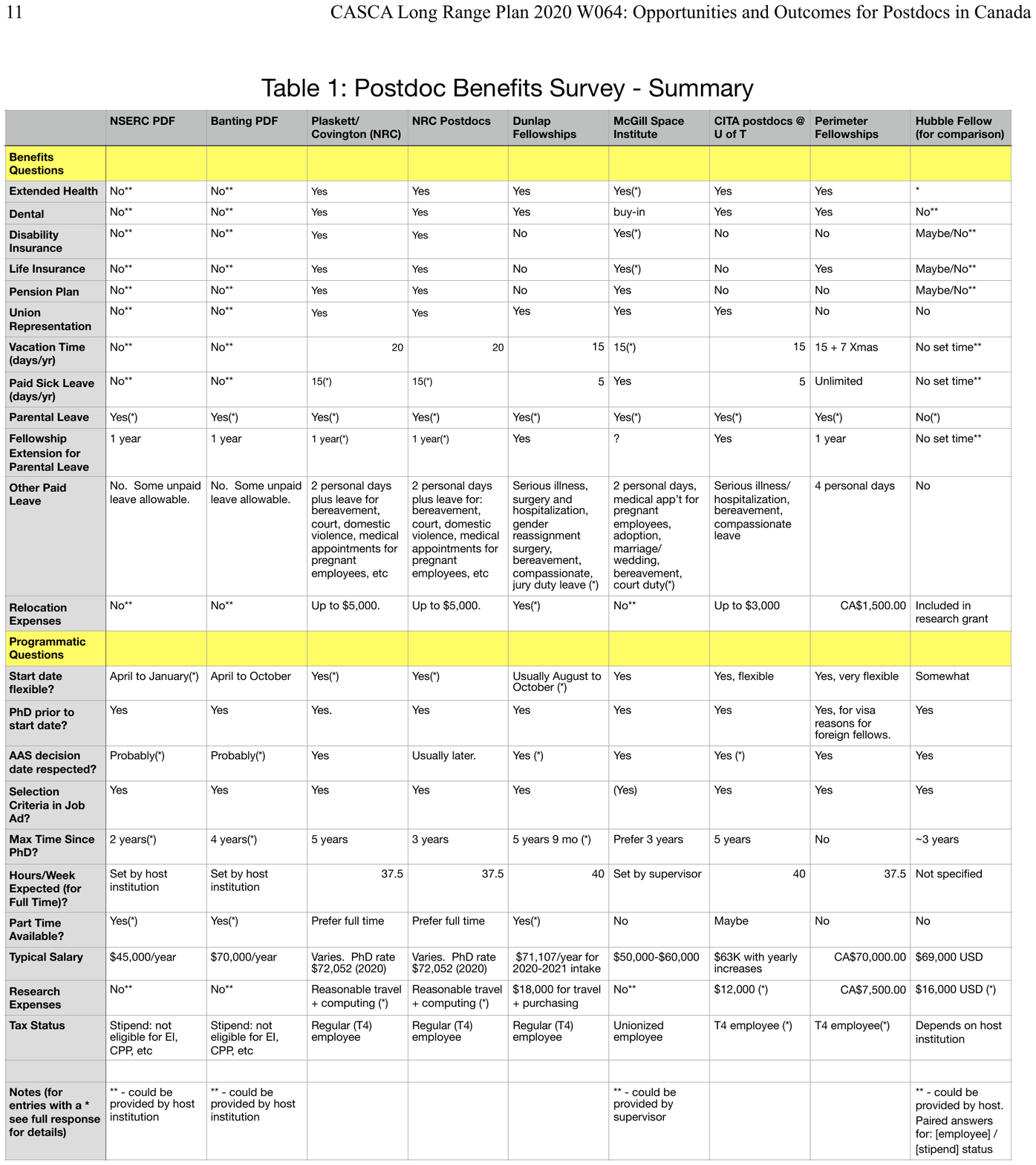}

\appendix

\section*{Appendix}

The following materials were not part of the official submission. Table 2 contains the full table of PDF benefits. We also include selected aggregated responses from a survey of PDFs in Canada (29 responses). This survey was conducted in July 2019.



\section*{Supplementary material}

## Table 2: Complete Postdoc Benefits Survey

| | NSERC PDF | Banting PDF | Plaskett/ Covington (NRC) | NRC Postdocs | Dunlap Fellowships | McGill Space Institute | CITA postdocs @ U of T | Perimeter Fellowships | Hubble Fellow (for comparison) |
|---|---|---|---|---|---|---|---|---|---|
| **Benefits Questions** | | | | | | | | | |
| **Extended Health** | None unless institutional host | None unless institutional host | Yes | Yes | Yes | Yes; parts are opt-out | Yes | Yes | Yes. Employees may have to cover part of the premium themselves. Up to $20,000/year covered for health premiums for family. |
| **Dental** | None unless institutional host | None unless institutional host | Yes | Yes | Yes | Yes, optional, you can buy in | Yes | Yes | No |
| **Disability Insurance** | None unless institutional host | None unless institutional host | Yes | Yes | No | Yes, may be allowed to opt-out | No | No | Employee - maybe; Stipend - no |
| **Life Insurance** | None unless institutional host | None unless institutional host | Yes | Yes | No | Yes, optional | No | Yes, 1 x annual salary | Employee - maybe; Stipend - no |
| **Pension Plan** | None unless institutional host | None unless institutional host | Yes | Yes | No | Yes, defined contribution | No | Nothing | Employee - maybe; Stipend - no |
| **Union Representation** | None unless institutional host | None unless institutional host | Yes, PIPSC | Yes, PIPSC | YES, CUPE 3902 Unit 5 | Yes | Yes | Not applicable | No |
| **Vacation Time** | None unless institutional host | None unless institutional host | Yes, 4 weeks per year | Yes, 4 weeks per year | 3 weeks per contract year | Yes, 15 working days for the first year, then increases by year | Yes - 15 days | 3 weeks plus 7 days at Christmas | Stipendees - several weeks suggested; no requirement as long as fellow maintains productivity. Employees - policy with host institution. |
| **Sick Leave** | None unless institutional host | None unless institutional host | Yes, up to 3 weeks per year (unspent time accumulates in following years) | Yes, up to 3 weeks per year (unspent time accumulates in following years) | 5 days per contract year | Yes | Yes - 5 days | Unlimited amount of paid sick days | Stipendees - no amount specified, suggestion to take as necessary. Employees - policy with host institution |
| **Parental Leave** | Yes, up to 1 year at full pay, subject to agency funding available. | Yes, up to 1 year at full pay, subject to agency funding available. | Maternity allowance: 17 weeks at 93% of regular pay (topping up EI). Parental allowance: 35-37 weeks at 93% of regular pay (topping up EI). Maximum parental allowance is 52 weeks for the couple. **This has been extended to 18 months (an additional 26 weeks). Employees can choose between option 1 (total 52 weeks), or option 2 (total of 78 weeks). The top up amount for the extended leave is approximately 64%. This generally adds up to the same amount the employee would receive during the regular leave, just extended over a longer length of time. | Maternity allowance: 17 weeks at 93% of regular pay (topping up EI). Parental allowance: 35-37 weeks at 93% of regular pay (topping up EI). Maximum parental allowance is 52 weeks for the couple. **This has been extended to 18 months (an additional 26 weeks). Employees can choose between option 1 (total 52 weeks), or option 2 (total of 78 weeks). The top up amount for the extended leave is approximately 64%. This generally adds up to the same amount the employee would receive during the regular leave, just extended over a longer length of time. | 17+35 weeks for birth parent, 37 weeks for non-birth parent; the University will pay the lesser of $800 dollars or 95% of salary during the one week waiting period for Employment Insurance benefits, provided that the employee applies for, and receives, Employment Insurance. For the next 15 weeks, or until the end of the appointment (whichever comes first), the University will pay the lesser of $400) or the difference between Employment Insurance benefits and 95% of the actual salary which the employee was receiving on the last day worked prior to the commencement of the maternity leave, provided that the employee applies for, and receives, Employment Insurance. | Paid parental leave (QPIP) PLUS top-up. QPIP details: Maternity leave: 18/15 weeks at 70%/75% of pay. Paternity leave: 5/3weeks at 70%/75% of pay. Parental leave (can be shared): 32weeks at (70% for first 7 weeks and 55% for remaining 25 weeks) OR 25 weeks at 75%. TOP-UP(union): 95% salary for first 20 weeks of maternity leave | Maternity leave: Up to 17 weeks. Top up to EI: First one week: $800 or 95% salary, whichever is less. Remaining 15 weeks: top up of $400 or 95% of salary, whichever is less  Parental leave: Up to 35 weeks (37 weeks for non-birth parent). Top-up to EI: First one week: $800 or 95% salary, whichever is less. Remaining 8 weeks: top up of $400 or 95% of salary, whichever is less Both: Leave and payments shall not continue beyond end date of contract | 7 weeks of maternity leave with top-up to EI benefits to 95% of salary. 35 weeks of parental leave unpaid. Postdoc not taking maternity leave would receive 2 weeks of parental leave paid at 100% of salary and 33 weeks of unpaid parental leave. | Not specified. General note that stipendees cannot take extended (paid) leaves, but may request no-cost extension following unpaid leave . Employees leave policy governed by host institution. |

| | NSERC PDF | Banting PDF | Plaskett/ Covington (NRC) | NRC Postdocs | Dunlap Fellowships | McGill Space Institute | CITA postdocs @ U of T | Perimeter Fellowships | Hubble Fellow (for comparison) |
|---|---|---|---|---|---|---|---|---|---|
| Fellowship Extension for Parental Leave | Yes, up to 1 additional year. | Yes, up to 1 additional year. | Yes, up to 1 additional year. Fellowship cannot extend beyond 5 years. | Yes, up to 1 additional year. Fellowship cannot extend beyond 5 years. | Yes | | Yes | Fellowship extension allowed for amount of leave taken or maximum of 1 year. | Possibly - as above |
| Other Paid Leave | No, but various reasons for unpaid leave are allowable. | No, but various reasons for unpaid leave are allowable. | Yes, including 2 personal days and as needed list including: bereavement, court leave (including jury duty), domestic violence, medical appointments for pregnant employees, and others. | Yes, including 2 personal days and as needed list including: bereavement, court leave (including jury duty), domestic violence, medical appointments for pregnant employees, and others. | Yes, e.g. serious illness, surgery & hospitalization: up to 2 months; Gender reassignment surgery leave: up to 2 months, Bereavement leave: 3 consecutive days (5 days if extensive travel required) per contract year; Compassionate leave: one week per contract year; Jury duty leave: 1 week. | Yes, up to 2 personal days and medical appointments for pregnant employees, adoption leave (10 weeks, topping up QPIP to full salary), marriage (5 days), wedding of close family member (1 day), bereavement leave (1-5 days depending on relationship), jury duty or other required court presence. | Yes - Serious illness/ hospitalization leave, bereavement leave, compassionate leave | 4 paid personal leave days per year | No |
| Relocation Expenses | None unless institutional host | None unless institutional host | Yes, up to $5,000. | Yes, up to $5,000. | Yes, amount depends on location | No, depends on the supervisor | Up to $3,000 | CA$1,500.00 | Yes, folded in with other research expenses |
| **Programmatic Questions** | | | | | | | | | |
| Start date flexible? | Yes. April to January following the offer, with deferment possible in some circumstances. | April to October of the year awarded. | Usually a multiple-month range acceptable | Usually a multiple-month range acceptable | YES, usually between August and October | Yes, the start date is flexible. | Yes, flexible | Yes, very flexible | Somewhat |
| PhD prior to start date? | Yes | Yes | Yes | Yes | Yes | Yes. Formal PhD requirements need to be met prior to starting because the university requires a PhD degree in order to register a postdoc. | Yes | Yes, for visa reasons for foreign fellows. | Yes |
| AAS decision date respected? | Probably. Offers nominally made at the end of January. | Yes. 2020 award offers nominally made mid-February | Yes | Yes, usually much later. | Yes for most cases | Yes | Acceptance date fluctuates, usually negotiated with candidate and extended to February 15 if requested | Yes | Yes |
| Selection Criteria in Job Ad? | Yes | Yes | Yes | Yes | Yes | No response given | Yes | Yes | Yes |
| Max Time Since PhD? | Generally 2 years. 3 years if industry employment post-PhD or career interruption. 6 years if primary caregiver to child within 2 years of PhD. | Yes. Limited to Sept 15, 2016 to Sept 30, 2020. Up to 2 extra years if career interruptions due to: parental leave, illness, health-related family responsibilities, mandatory military service, disruptions due to war, civil conflicts and/or natural disasters in the country of residence | Yes, within the last 5 years. | Yes, within the last 3 years. | Yes, PhD granted in the last 5 years 9 months barring extenuating circumstances | No set maximum, but preference is given to applicants within the last 3 years preferred of their PhDs | Within 5 years of completing PhD as required by university regulations | No | Yes, PhD granted since January 1, 2017 |
| Hours/Week Expected? | Set by host institution | Set by host institution | 37.5 | 37.5 | 40 hours/week if full-time | Depends on the supervisor | 40 | 37.5 | Not specified |
| Part Time Available? | Yes, for parental or medical reasons and/or family responsibilities. Needs institutional and NSERC approval. Awards pro-rated in value and duration in this instance. | Yes, for parental or medical reasons and/or family responsibilities. Needs institutional and NSERC approval. Awards pro-rated in value and duration in this instance. | Generally, we would want full time. | Generally, we would want full time. | Yes subject to immigration regulations for non-Canadian | No, minimum 6-months appointment to be registered as a postdoc | Has not been requested yet | No | No |
| Typical Salary | $45,000/year | $70,000/year | Varies with experience. PhD recruiting rate for 2020 $72,052 (from Plaskett ad) | Varies with experience. PhD recruiting rate for 2020 $72,052 (from Plaskett ad) | $71,107/year for 2020-2021 intake | Min. $34,100 per the collective agreement. Typical salary range $50k - $60k | $63K with yearly increases | CA$70,000.00 | $69,000 USD |

|  | NSERC PDF | Banting PDF | Plaskett/ Covington (NRC) | NRC Postdocs | Dunlap Fellowships | McGill Space Institute | CITA postdocs @ U of T | Perimeter Fellowships | Hubble Fellow (for comparison) |
|---|---|---|---|---|---|---|---|---|---|
| **Research Expenses** | Only if host institution provides | Only if host institution provides | Shared group travel budget, with postdocs prioritized. Computing equipment provided as needed. | Shared group travel budget, with postdocs prioritized. Computing equipment provided as needed. | $18,000 for travel + purchasing | Depends on the arrangement reached with the PI. There is no travel expenses attached to the MSI fellowship as of right now. | $12K research fund that they can use at their own discretion with longer trips requiring prior approval from director | $7,500 annually | $16,000 USD (includes relocation at start, if desired). |
| **Tax Status** | Stipend, not salary. Not eligible for EI, CPP, etc | Stipend, not salary. Not eligible for EI, CPP, etc | Regular employees - get T4 | Regular employees - get T4 | Employees - get T4 | Postdocs funded by MSI fellowships are treated by McGill as unionized employees | CITA postdocs are unionized employees, there are legislated tax deductions, including CPP, but no pension plans, as per collective agreement | Treated as employees who contribute to CPP, EI, Fed & Prov taxes | Employee or stipend depending on host institution |
| **Other Notes** | Survey filled in by authors using information online only | Survey filled in by authors using information online only |  |  |  | Fellows are partially funded by MSI and partially by supervisor | Other CITA Fellows (e.g., CITA National Fellows) are only partially supported by CITA and benefits/rules mostly set by host. |  | Survey filled in by authors using information online only. |

# Selected Responses from Postdoc Survey

Note: Some responses are omitted or aggregated to avoid reporting of small bins.

## Demographics of respondents

Fig A1: How many years of total postdoc experience do you have?

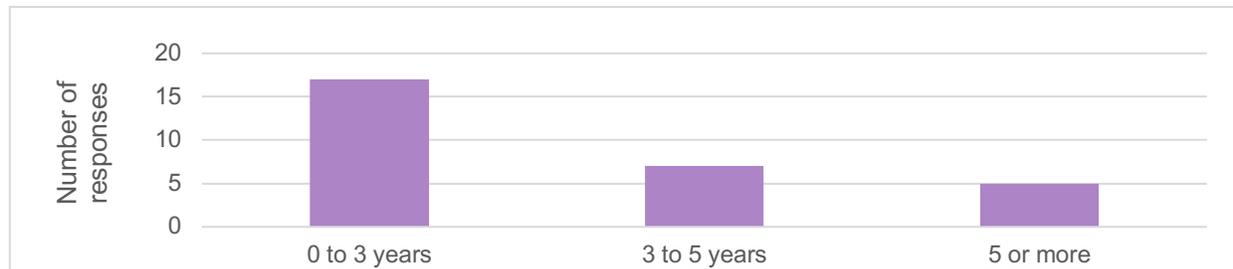

Fig A2: How many years is your current postdoc position?

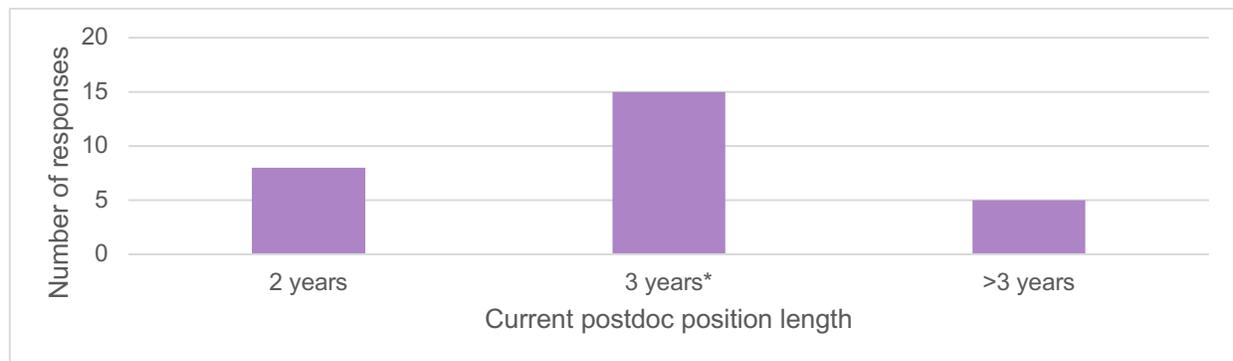

* The "3 years" category includes responses of "2+1 years"

Fig A3: Including your current postdoc, how many postdoc positions have you held?

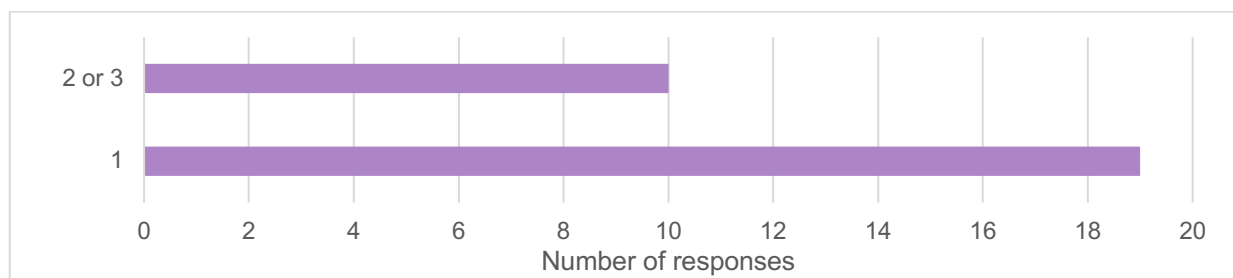

## Postdoc compensation and benefits

Fig A4: What is your annual salary as a postdoc (based on a full-time appointment)?
*Note: Some responses not included due to small number of responses within that cell*

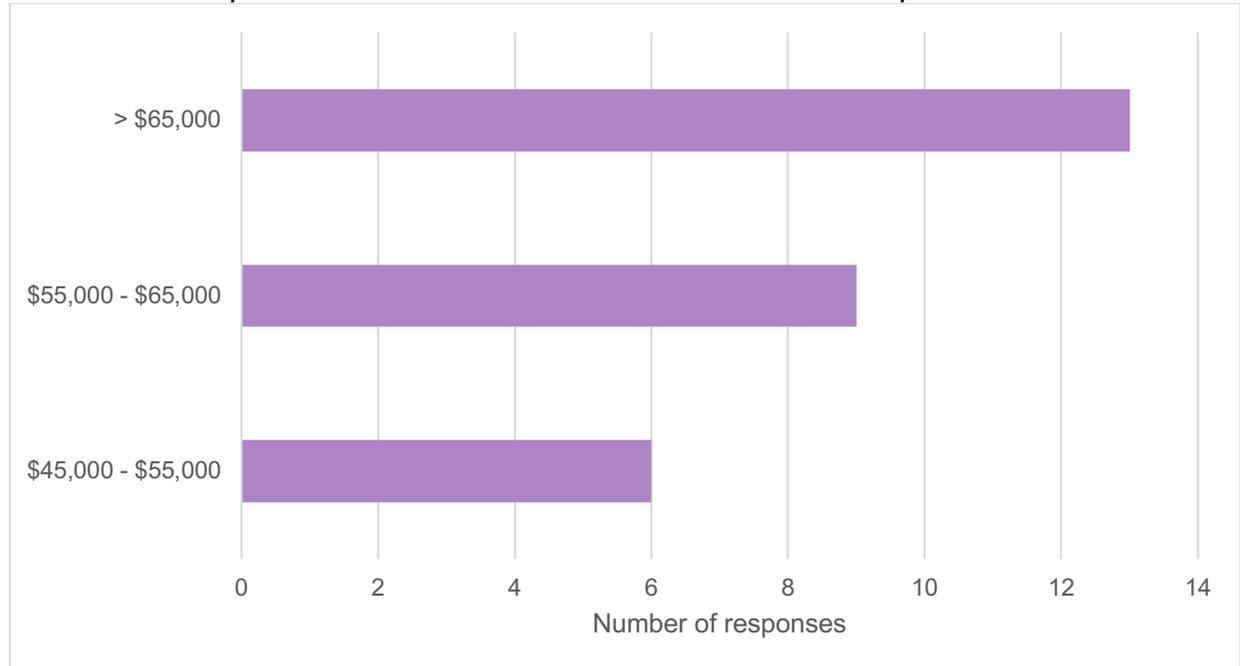

Fig A5: For the respondents who moved geographical location for their postdoc: Were you provided with an adequate moving allowance?

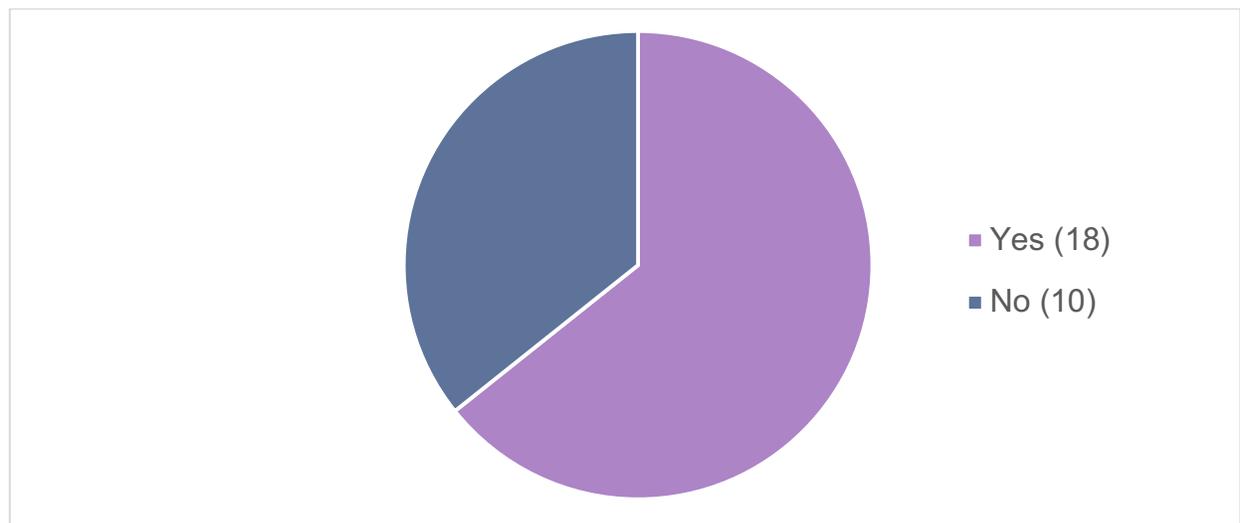

Fig A6: Which benefits are you entitled to in your postdoc?

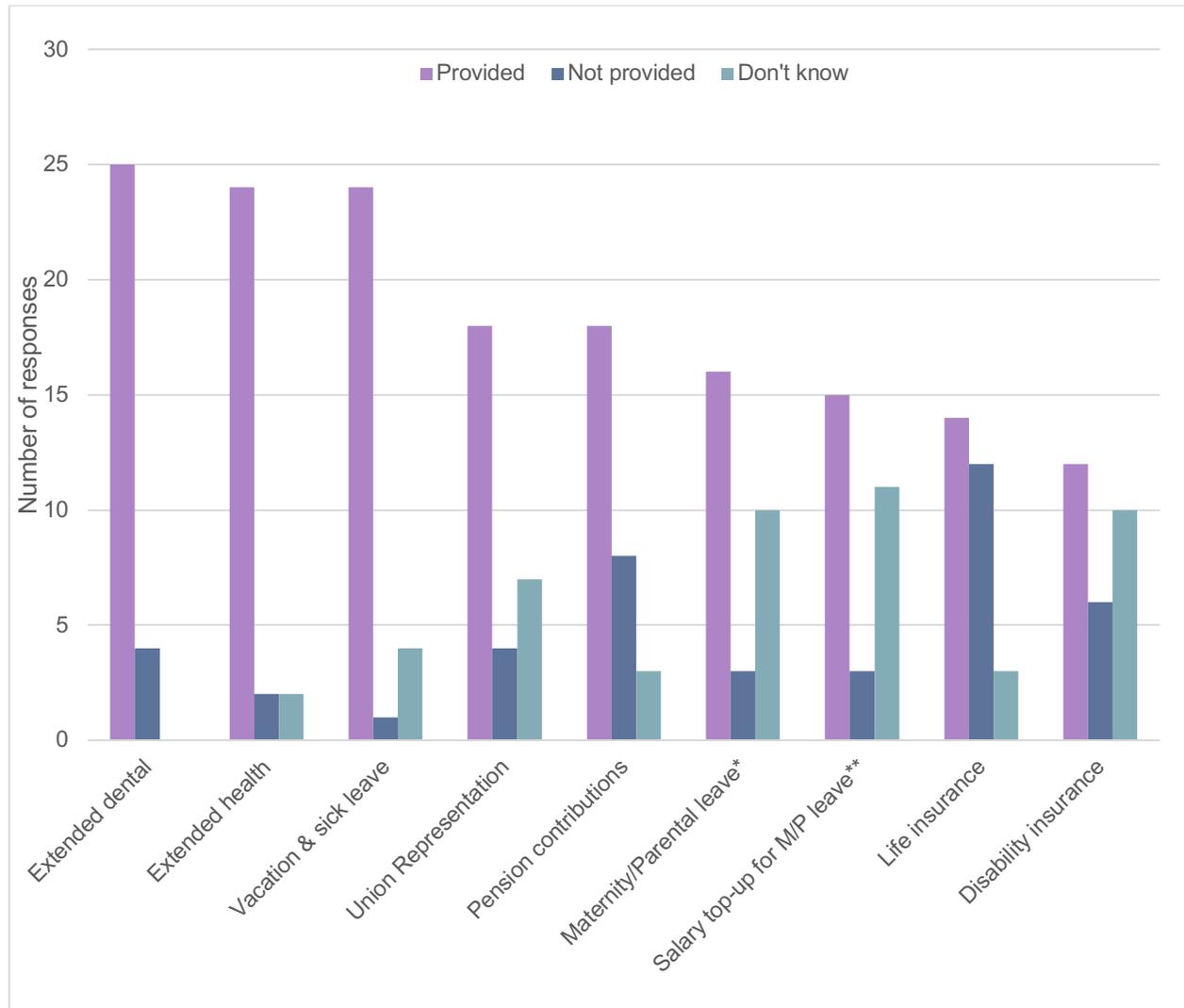

Notes
* Maternity and parental leave without pay
** Any amount of salary top-up for maternity/parental leave for any number of weeks

## Postdoc work responsibilities and resources

Fig A7: Do you do any of the following non-research tasks as part of your postdoc?

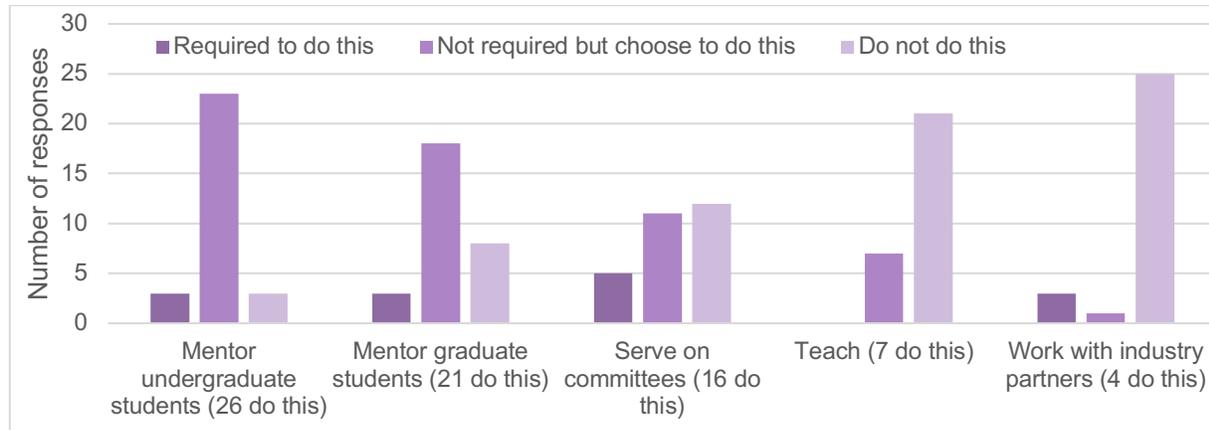

Fig A8: How often do you work more than your institution's full-time equivalent hours (typically 35-40 hours per week)?

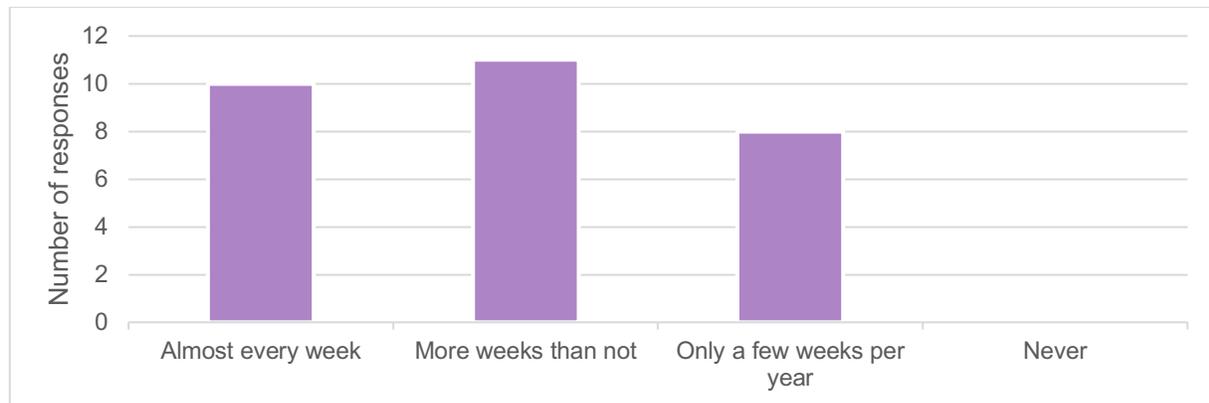

Fig A9: Have you ever been required to pay work/conference travel costs yourself as part of your postdoc?

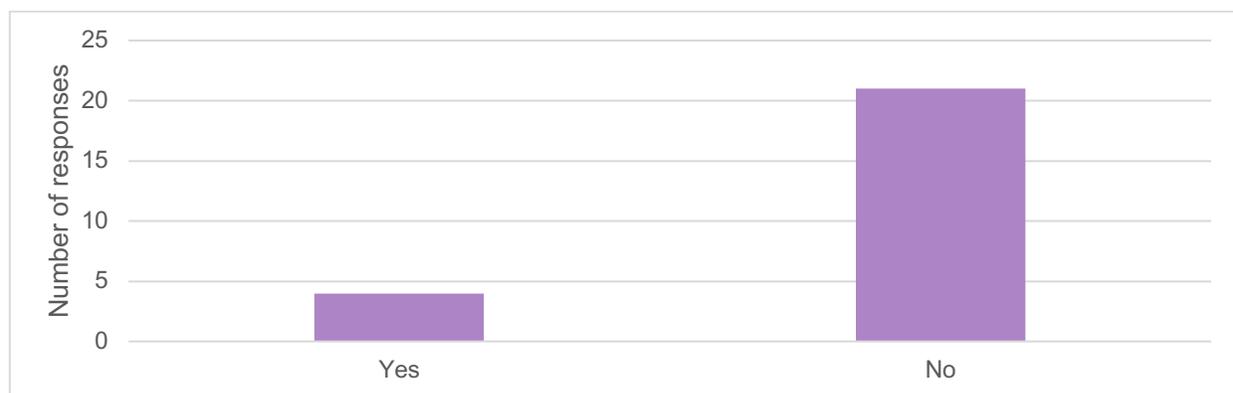

## Postdoc international and family status

Fig A10: Do you have any children or underage dependents for which you are a primary caregiver?

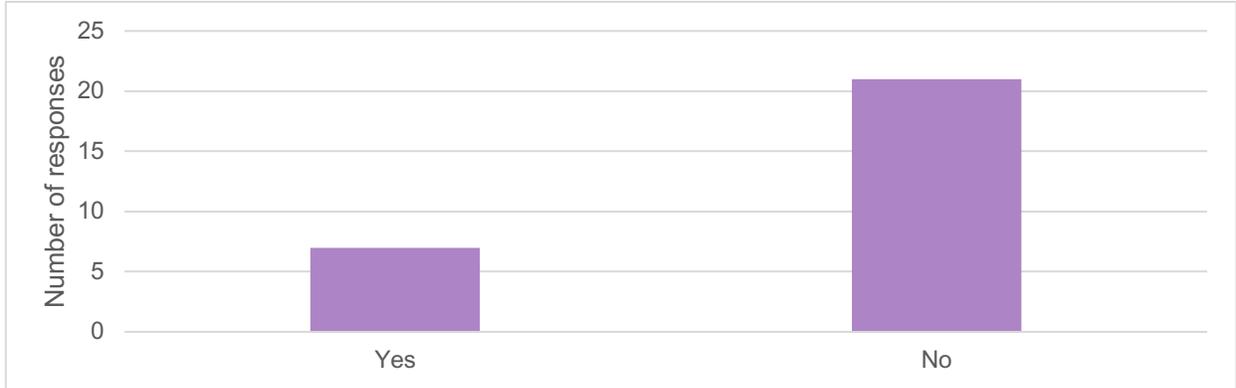

Fig A11: Are you in Canada under a work visa and/or work permit?

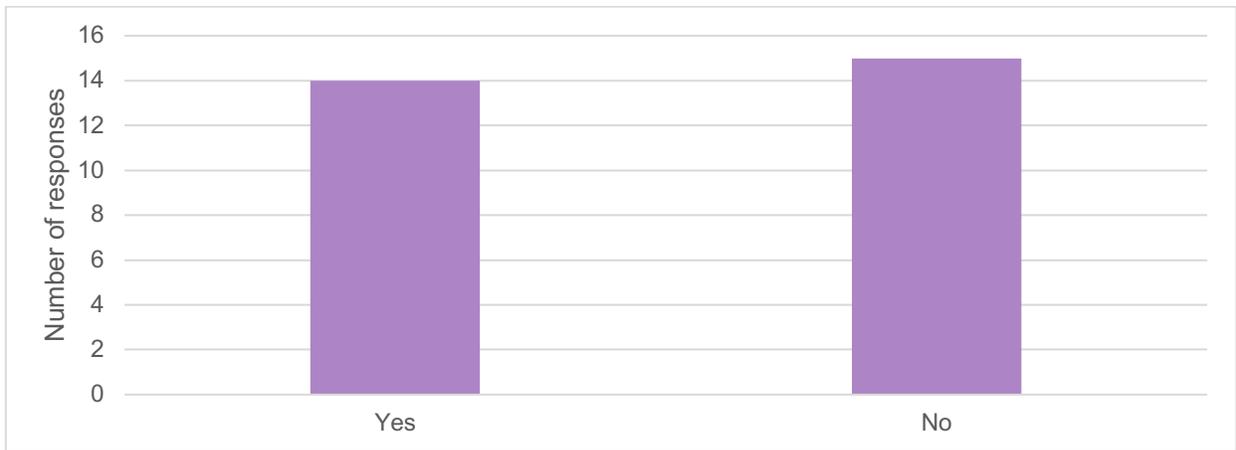

## Postdoc Outcomes

Fig A12: Upon the end of your postdoc, what type of employment do you hope for? (choose all that apply)

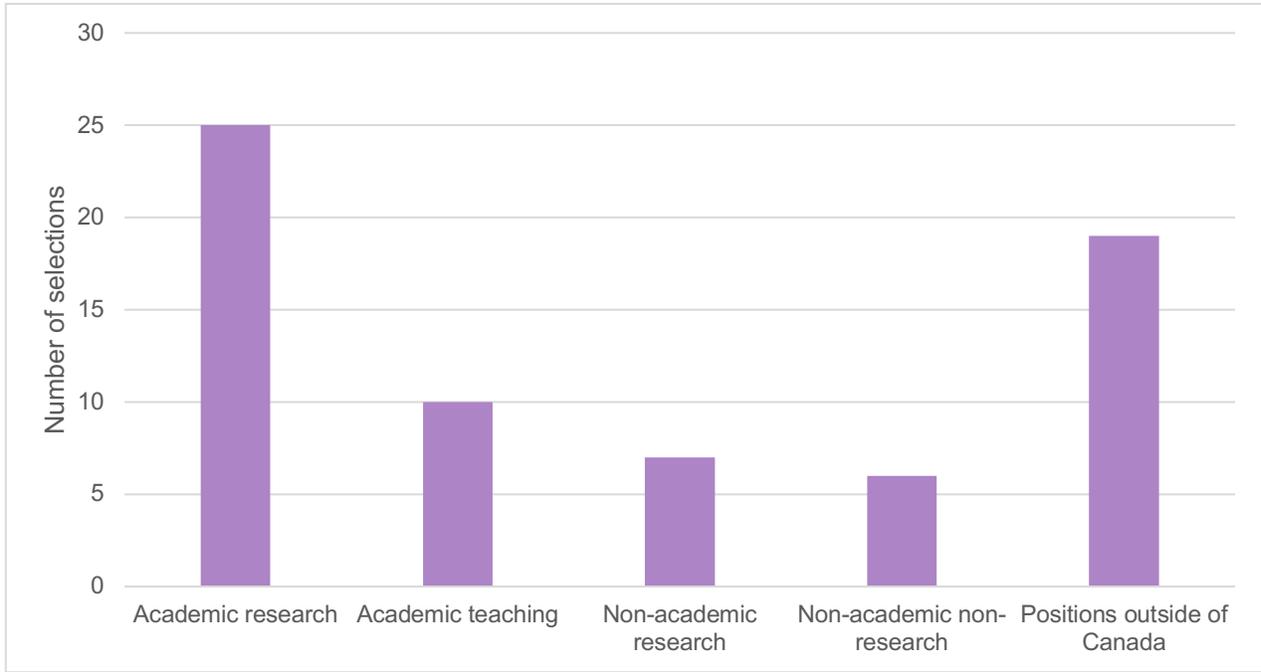

Fig A13: Do you believe that you are provided with enough resources to help you gain employment (either academic or otherwise) after your contract ends?

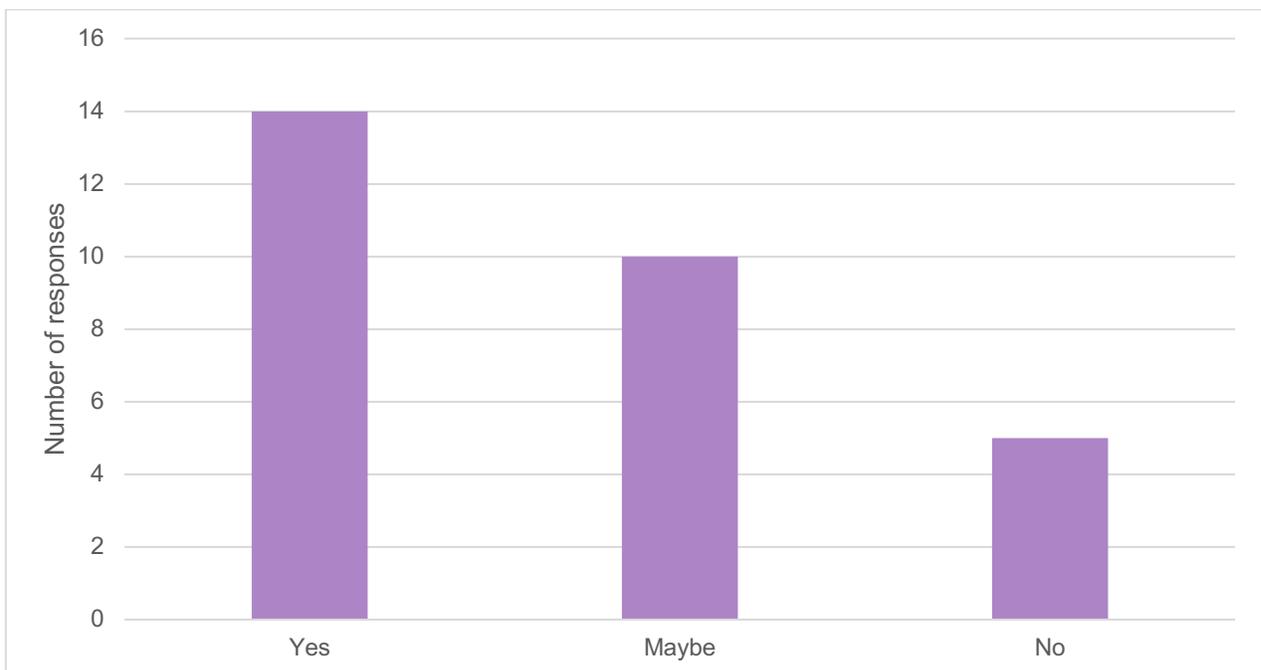

## Open-ended questions

Here, we present summaries of responses to three open-ended questions.

**What resources would help you better prepare for employment upon the end of your post?**

Training-related (4 responses):
- Broader training of skills such as teaching, project management, data science
- Discussions of how postdocs develop and apply professional skills
- Workshops on marketing postdoc skills to industry

Careers-related (4 responses):
- Better connections between current and future employers, e.g. Universities taking a role in promoting their PDFs to local employers, job fairs, etc.
- Networking events with non-academic employers
- Discussions/workshops on transition from academia to industry

Mentorship-related (2 responses):
- Formal mentoring with both faculty members and non-academic mentor, especially those who were recently successful in the job market

Teaching experience opportunities
Longer postdocs and more funding for postdocs


**What are the important issues to highlight during this LRP process?**
*Note: This question had three checkboxes: "More jobs", "More funding" and "Professional Development", with space to write in comments. Here is a summary of the responses.*

"More jobs": Marked by 18/27 respondents. In the open answer part, respondents emphasized the need for more stability (longer postdocs lengths) and more permanent positions instead of more term positions.
"More funding": Marked by 17/27 respondents.
"Professional Development": Marked by 15/27 respondents.

**Stories, comments and anecdotes submitted by respondents**
*Note: Some comments abridged or edited for brevity and anonymity.*

Comment #1: I went on to a second postdoc [outside of Canada], and am now about to start a faculty position [outside of Canada]. My first postdoc in Canada was key to me feeling welcomed and part of the world astronomy community—and it was great fun.

Comment #2: Canadian astronomers nearly always have to leave the country to find postdoc or permanent positions. It is generally hard to move so far, so often. It is harder

when you have a spouse and/or dependents. Losing your support network when you have children is very challenging. In addition, some countries make it very difficult for a spouse to obtain a work permit, and a single postdoc salary is not usually enough to support a family (especially if you still carry student loan debt). I ended up in debt after my first postdoc in the US.

Comment #3: Towards the end of my current contract, I began to apply for postdoc positions within Canada, as well as other countries. […] I never heard a single thing back - not even a rejection email. The only reason I found out that I had not been considered was through the Astro Rumour Mill, which is simply unacceptable.

Whilst I understand that large fellowships like these [...] attract a huge number (100s) of applicants, it is incredibly demoralizing to not even receive a response back after all the effort put into an application. All it takes is a 1-2 line email saying "Dear X, we have reviewed your application for Y and unfortunately, at this time, we will not be able to offer you a position this year." This would take a day of somebody's time at the very most. There are very limited postdoctoral positions in Canada and people put an awful lot of time and effort into their applications in order to stand the best chance of getting a job. To not hear a response back, nothing, not even a 1 line rejection email, makes you think that people at these institutes haven't even read your application. Please share this story, because people, especially the hiring committees at larger institutes need to hear this, and they need to know how it feels to be completely ignored. The postdoc job hunt is cutthroat and brutal enough without institutes being too lazy to send out formal rejection letters.

Comment #4: I'd like to see Canadian conferences and organizations become even more understanding towards the fact that many of postdocs here are speaking English as a second language. This can improve outcomes. What are some of the ways that we as a community can become more understanding and accommodating? Science is international.

Comment #5: I actually recently finished my postdoc and successfully got a faculty position outside of Canada, but filled out the form since I thought the data would be useful to you. Even though I had a fairly prestigious postdoc, my sense is the pay in Canada is awful; at $55k I made less than my previous postdoc in the US. While I was treated fine, postdoc seemed like an afterthought at my institution. There weren't very many of us, and as a 2yr (with potential for a 3rd from other funding) contract, there was very little sense of community for those at this career stage. It felt isolating.



\end{document}